\documentclass[%
reprint, 
superscriptaddress, 
amsmath,amssymb,
]{revtex4-2}

\usepackage{graphicx}
\usepackage{epstopdf}
\usepackage{subcaption}
\usepackage{dcolumn}
\usepackage{bm}
\usepackage[colorlinks, linkcolor=red, anchorcolor=blue, citecolor=green]{hyperref}
\usepackage{dcolumn}
\usepackage{soul, color, xcolor} 
\usepackage{amsmath}%
\usepackage{float}
\usepackage{pifont} 

\usepackage{braket}
\usepackage{array}
\usepackage{booktabs}             
\usepackage{threeparttable}       
\usepackage{multirow}     
\usepackage{diagbox}    
\usepackage{array}      
\usepackage{caption}    
\usepackage{ragged2e}
\usepackage{ulem}
\usepackage{xspace}
\usepackage{times}

\newcommand{\addressa}{\affiliation{Key Laboratory of Quantum Information, University of Science and Technology of China, Hefei, Anhui, 230026, China}}
\newcommand{\addressb}{\affiliation{CAS Center For Excellence in Quantum Information and Quantum Physics, University of Science and Technology of China, Hefei, Anhui, 230026, China}}
\newcommand{\addressc}{\affiliation{Institute of Artificial Intelligence, Hefei Comprehensive National Science Center, Hefei, Anhui, 230088, China}}
\newcommand{\addressd}{\affiliation{Institute of the Advanced Technology, University of Science and Technology of China, Hefei, Anhui, 230088, China}}
\newcommand{\addresse}{\affiliation{Origin Quantum, Hefei, Anhui 230026, China}}
\newcommand{\addressf}{\affiliation{Suzhou Institute for Advanced Research, University of Science and Technology of China, Suzhou, Jiangsu 215123, China}}

\begin{document}

\preprint{APS/123-QED}

\title{Realizing Scalable Conditional Operations through Auxiliary Energy Levels }

\author{Sheng Zhang}
\addressa\addressb\addressf
\author{Peng Duan}
\email{pengduan@ustc.edu.cn}
\addressa\addressb

\author{Yun-Jie Wang}
\addressd

\author{Tian-Le Wang}
\author{Peng Wang}
\author{Ren-Ze Zhao}
\author{Xiao-Yan Yang}
\author{Ze-An Zhao}
\author{Hai-Feng Zhang}
\addressa\addressb
\author{Liang-Liang Guo}
\author{Yong Chen}
\author{Lei Du}
\author{Hao-Ran Tao}
\addresse
\author{Zhi-Fei Li}
\author{Yuan Wu}
\addressa\addressb

\author{Zhi-Long Jia}
\author{Wei-Cheng Kong}
\addresse

\author{Zhao-Yun Chen}
\addressc

\author{Zhuo-Zhi Zhang}
\author{Xiang-Xiang Song}
\addressa\addressb\addressf
\author{Yu-Chun Wu}
\addressa\addressb

\author{Guo-Ping Guo}
\email{gpguo@ustc.edu.cn}
\addressa\addressb\addresse


\date{\today}

\begin{abstract}
In the noisy intermediate-scale quantum (NISQ) era, flexible quantum operations are essential for advancing large-scale quantum computing, as they enable shorter circuits that mitigate decoherence and reduce gate errors. However, the complex control of quantum interactions poses significant experimental challenges that limit scalability.
Here, we propose a transition composite gate scheme based on transition pathway engineering, which digitally implements conditional operations with reduced complexity by leveraging auxiliary energy levels. 
Experimentally, we demonstrate the controlled-unitary (CU) family and its applications. In entangled state preparation, our CU gate reduces the circuit depth for three-qubit Greenberger–Horne–Zeilinger (GHZ) and W states by approximately 40–44\% compared to circuits using only CZ gates, leading to fidelity improvements of 1.5\% and 4.2\%, respectively.
Furthermore, with a 72\% reduction in circuit depth, we successfully implement a quantum comparator—a fundamental building block for quantum algorithms requiring conditional logic, which has remained experimentally challenging due to its inherent circuit complexity.
These results demonstrate the scalability and practicality of our scheme, laying a solid foundation for the implementation of large-scale quantum algorithms in future quantum processors.

\end{abstract}
\maketitle


\section{\label{sec:level1}Introduction}

Superconducting quantum computers have progressed into noisy intermediate-scale quantum (NISQ) devices with significant potential for demonstrating quantum advantage\cite{preskill2018quantum,kjaergaard2020superconducting,arute2019quantum,gao2025establishing}. However, the experimental realization of large-scale quantum algorithms is hindered by limited circuit depths due to the cumulative effects of gate errors and decoherence\cite{fowler2012surface,gonzalez2021digital,arute2019quantum}. A pivotal strategy to overcome these challenges is developing quantum operations with enhanced operational degrees of freedom. Such advancements can effectively reduce the required circuit depth, thereby enabling the implementation of quantum algorithms on larger-scale systems\cite{sim2019expressibility}.

The significance of more expressive quantum operations drives extensive research efforts. 
Highly expressive operations can effectively parameterize a vast and diverse set of unitary transformations, thereby providing a powerful representational capacity for approximating complex quantum processes~\cite{peterson2020fixed,abrams2020implementation}.
These operations are primarily categorized into native and composite operations. Native operations depend on engineering novel interactions to implement flexible operations, such as arbitrary two-qubit and multi-qubit gates~\cite{roy2020programmable,kim2022high,glaser2023controlled,chen2025efficient}. However, these approaches face challenges in multiple parameter optimization, increased system initialization time cost, and complex crosstalk corrections, thereby limiting their scalability \cite{caldwell2018parametrically,hong2020demonstration,sete2021parametric}.
By contrast, composite operations adopt a more straightforward strategy by assembling well-established gate components via Clifford decompositions, with the fSim and XY-family gates serving as typical examples~\cite{foxen2020demonstrating,abrams2020implementation}. Nevertheless, the limited degrees of freedom in existing components, like controlled phase (CZ) and iSWAP gates, can still lead to increased circuit depth~\cite{li2019realisation,rol2019fast,negirneac2021high}. Moreover, incorporating various components often exacerbates frequency collision issues, further complicating the allocation of qubit working frequencies~\cite{brink2018device,osman2023mitigation}. Thus, while both strategies introduce new schemes at the Hamiltonian and circuit-design levels to achieve more expressive gate sets, they still require careful trade-offs between complex interaction design and limited circuit depth.	

In this work, we propose the Transition Composite Gate (TCG) scheme, which facilitates the realization of highly expressive gate sets with reduced circuit depth.
Instead of auxiliary qubits, the TCG scheme temporarily utilizes the inherent non-computational space in superconducting quantum systems as auxiliary energy levels. Through precise engineering of transition pathways between computational and non-computational subspaces, it digitally constructs the expected conditional operations.
As a demonstration, we experimentally implement a controlled-unitary (CU) operation on a superconducting platform using the TCG scheme, which comprises two non-adiabatic square-root controlled phase ($\sqrt{\rm CZ}$) gates and a layer of single-qubit gates. Quantum process tomography (QPT) confirms the CU family, achieved by tuning single-qubit gate parameters, with fidelities ranging from 95.2\% to 99.0\%~\cite{pedersen2007fidelity,bialczak2010quantum,magesan2011gate}. Furthermore, the preparation of three-qubit Greenberger–Horne–Zeilinger (GHZ) and W states achieves circuit depth reductions of 40.0\% and 44.0\%, respectively, along with improved fidelities. Finally, the implementation of a quantum comparator with significantly reduced circuit depth demonstrates the application potential of the TCG scheme. Our work provides a critical high-expressivity operation scheme for large-scale quantum computing and holds promise for applications in diverse quantum algorithms and simulations.

\section{\label{sec:level2}Methods}
\subsection{\label{sec:leve21}Theoretical scheme}

To satisfy addressing requirements, quantum units in solid-state quantum computing are typically designed with non-uniform energy level structures. For example, the transmon qubit employed in this study can be treated as a qutrit~\cite{koch2007charge,barends2013coherent}, featuring an anharmonic three-level structure, as illustrated in Fig.~\ref{fig principle}(a). In qubit systems, the computational subspace is spanned by the states $|0\rangle$ and $|1\rangle$ ($\rm S_c$), while state $|2\rangle$ and higher corresponding to the non-computational space ($\rm S_{uc}$). 

While the occupation in $\rm S_{uc}$ is conventionally regarded as leakage in qubit systems and often suppressed in previous studies, it is exploited as a computational basis in qutrit-based system~\cite{chen2016measuring,barends2019diabatic,hyyppa2024reducing}. 
Although ternary logic based on qutrits presents challenges in control complexity and algorithm compatibility, it offers noval insights for quantum gate design.
In particular, the temporary exploitation of $\rm S_{uc}$ facilitates the operation construction in qubit systems~\cite{klimov2003qutrit,goss2022high,baekkegaard2019realization}. For instance, non-adiabatic CZ gates can accumulate a controlled phase via swap operations between the $|11\rangle$ and $|20\rangle$ states~\cite{li2019realisation,rol2019fast,negirneac2021high}, and this mechanism is employed in QuAnd gates to realize multi-controlled operations.~\cite{chu2023scalable}. 
Building upon this, by strategically designing and utilizing these auxiliary states as intermediate transition pathways, we can achieve conditional operations with reduced circuit depth compared to conventional Clifford decompositions, which we call the Transition Composite Gate (TCG) scheme, as shown in Fig.~\ref{fig principle}(b).
Specifically, our implementation constructs a gate sequence to realize the desired conditional operation through non-computational subspace, with the general framework provided in Appendix~\ref{Theoretical Framework}. 
We employed traditional Pauli gates along with the $\rm X^{12}$ gate between states $|1\rangle$ and $|2\rangle$, and the  $\mathrm{iSWAP}_{20\leftrightarrow 11}$ gate between states $|20\rangle$ and $|11\rangle$ (also known as $\sqrt{\rm CZ}$ gate)~\cite{wang2021optimal,chen2023transmon,chu2023scalable}.
In the following sections and Appendix~\ref{Examples}, we demonstrate several widely used controlled operations constructed from these well-established components.

It is worth noting that previous studies have indicated that utilizing high-energy levels introduces additional decoherence channels~\cite{morvan2021qutrit}. Therefore, a trade-off may be required between the occupation time of high-energy levels and the total operation duration. Additionally, the active transition into non-computational subspace and imperfect control introduce a higher risk of leakage. However, these technical challenges can be addressed through continuous improvements in qubit performance and optimization of control techniques.
Therefore, in this work, we focus on verifying the feasibility of the TCG scheme and assessing its advantages in quantum circuit implementation.

\subsection{\label{sec:leve22} Controlled-unitary gate}

\begin{figure}[ht]
\centering
\includegraphics[width=1.0\linewidth]{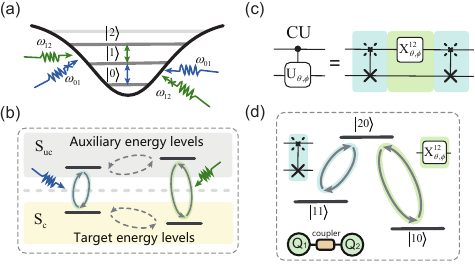}
\caption{\justifying
(a) Energy diagram of a typical transmon as a multilevel system. Microwaves of different frequencies are represented by arrows of various colors, inducing transitions only between corresponding energy levels. Non-adiabatic two-qubit operations can similarly bring specific transitions due to different implementation conditions, such as distinct resonance frequencies. These implie that the transition operations between $\rm S_c$ and $\rm S_{uc}$ are inherently independent due to the non-uniform energy level structures.
(b) The principle diagram of the TCG scheme. Instead of auxiliary qubits, we employ $\rm S_{uc}$ as auxiliary energy levels, while the operations within the $\rm S_{uc}$ and $\rm S_{c}$ are mutually independent. This allows one to reconstruct conditional operations by deliberately transferring the quantum system into the $\rm S_{uc}$, allocating independent degrees of freedom there, and subsequently returning to the $\rm S_{c}$. This procedure is referred to as transition pathway engineering.
(c) Illustration of the composition of CU gate under TCG scheme.
$\rm X^{12}$ gate is acted on the high-frequency qubit to concatenate transition channel.
(d) The state evolution diagram corresponding to (c).	$\sqrt{\rm CZ}$ and $\rm X^{12}$ gates independently induce interaction between states $|11\rangle$-$|20\rangle$ and states $|20\rangle$-$|10\rangle$.
}
\label{fig principle}
\end{figure}

The flexible and efficient CU gate plays a crucial role in quantum circuits~\cite{tuanuasescu2022distribution}. However, native CU gates are constrained by complex physical implementations, while digital schemes tend to result in deeper circuit depths.
Here, the TCG scheme provides a straightforward construction of the CU gate, wherein two $\sqrt{\rm CZ}$ gates sandwich an $\rm X^{12}$ gate, as illustrated in Fig.~\ref{fig principle}(c).
The states $|11\rangle$ and $|10\rangle$ are indirectly connected through $\sqrt{\rm CZ}$ and $\rm X^{12}$ gate, with state $|20\rangle$ serving as the ``bridge'' for commuting within ${\rm S}_{\rm c}$.
Figure~\ref{fig principle}(d) illustrates specific transition pathways of our CU operation, and the mathematical description within ${\rm S}_{\rm c}$ is as follows:
\begin{gather}
\rm {CU}= \sqrt{\rm CZ} \cdot \mathit{I}_0 \otimes X_{\theta,\phi}^{12} \cdot \sqrt{\rm CZ} \\
=
\begin{pmatrix}
1 & 0 & 0 & 0 	\\
0 & 1 & 0 & 0 	\\
0 & 0 & \cos(\frac{\theta}{2}) & -ie^{-i\phi}\sin(\frac{\theta}{2})	\\
0 & 0 & -ie^{i\phi}\sin(\frac{\theta}{2}) & -\cos(\frac{\theta}{2}) 
\end{pmatrix}, 
\end{gather}
where $\theta$ and $\phi$ represent the Rabi angle and phase of the $\rm X^{12}$ gate, respectively. $I_0$ is the identity operation.	Here, $\sqrt{\rm CZ}$ operation acts as the adjustable component involving complex processes, while the $\rm X^{12}$ gate is simple to control as a single-qubit operation. Hence, we are inclined to use microwaves to control the latter. Note that the control of $\theta$ and $\phi$ is associated separately with amplitude ($A$) and phase ($\varphi$) of microwave pulse. This implies that we can achieve precise and independent control under well-established techniques. 
Moreover, simply adding a single-qubit phase gate can further provide additional control over the phase.	Our focus lies in controlling $\theta$ and $\phi$ in this work, and we give the complete derivation process in Appendix~\ref{Examples}.

The CU operation under the short-path TCG scheme (SPTCG) further reduces the components required, achieving a unidirectional state transition process, as shown in Fig.~\ref{fig entaglement}(a). This short-path CU operation (SPCU) can be described within the ${\rm S}_{\rm c}$ as:
\begin{gather}
{\rm SPCU} = I_0 \otimes {\rm X}_{\theta,\phi}^{12} \cdot \sqrt{\rm CZ} \\
= 
\begin{pmatrix}
    1 & 0 & 0 & 0 	\\
    0 & 1 & 0 & 0 	\\
    0 & 0 & \cos(\frac{\theta}{2}) & 	-ie^{-i\phi}\sin(\frac{\theta}{2})	\\
    0 & 0 & 0 & 0 
\end{pmatrix}.
\end{gather}
The absence of the first $\sqrt{\rm CZ}$ operation still permits a transition from state $|10\rangle$ to $|11\rangle$, while the absence of the second $\sqrt{\rm CZ}$ operation corresponds to the reverse process. Although this omission renders the composite operation irreversible and unsuitable as a quantum gate, it retains its capability for state preparation while reducing the overall time cost. We provide more details in Appendix~\ref{Theoretical Framework}.

In general, compared to conventional Clifford decompositions (consisting of two CNOT gates and three single-qubit gates), our approach achieves a 40\% reduction in circuit depth along with lower time costs. This advantage becomes increasingly significant in the construction of controlled multi-qubit gates, thereby alleviating error accumulation and decoherence in large-scale quantum computing.

\section{\label{sec:level3}Results}
\subsection{\label{sec:level31} Control rotation and Tunable phase}

\begin{figure}
\centering
\includegraphics[width=1.0\linewidth]{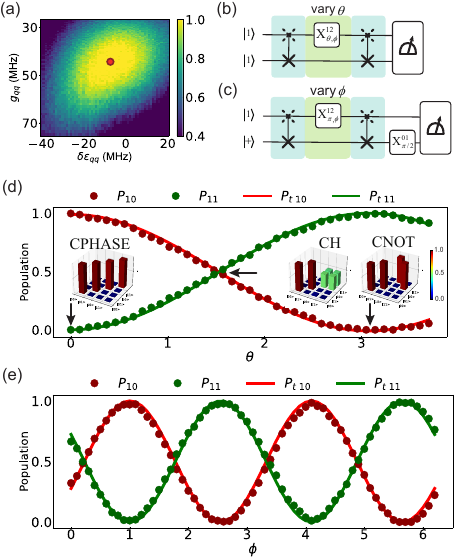}
\caption{\justifying
(a) Landscapes of $P_{20}$ as a function of detune ($\delta \varepsilon_{qq}=\varepsilon_{|11\rangle}-\varepsilon_{|20\rangle}$) and effective coupling strength ($g_{qq}$).
The operation time is 30 ns.
(b) and (c) show the schematic illustration of the verification setup for the controlled rotation and tunable phase. 
(d) Experimental results graph under (c). 
Dots represent experimental data with 5000 shots, while lines depict theoretical predictions.
Cphase, CH, and CNOT gates can be implemented at different $\theta$, respectively.
Theoretical (gray) and experimental (colored) truth tables for these gates are shown in the small figure.
(e) The experimental results graph under (d).
There exist relative values between the $\phi$ and $\varphi$, originating from the relative phase of $\rm X^{12}$ and $\rm X^{01}$ and the two-qubit interaction. More details and experimental verifications are provided in Appendix~\ref{Additional Verification}.
}
\label{fig tunable}
\end{figure}

First, we carefully adjusted the waveform pulse of the $\sqrt{\rm CZ}$ component to achieve the full exchange between the states $|11\rangle$ and $|20\rangle$.
Specifically, we determined parameters of the coupler and qubits by a peak-finding algorithm, as shown in Fig.~\ref{fig tunable}(a). We also implemented the Derivative Reduction by Adiabatic Gate (DRAG) scheme on the $\rm X^{12}$ gate to minimize leakage and phase errors from other energy levels~\cite{motzoi2009simple}.
Next, we conducted a preliminary characterization of the CU operation using the setups depicted in Figs.~\ref{fig tunable}(b) and (c). 
We varied the $\theta$ in the CU operation and observed the final states when the qubit pair was initially in the state $|11\rangle$ to verify the tunable rotation. The results presented in Fig.~\ref{fig tunable}(d) show that the population of $|10\rangle$ ($P_{10}$, green points) varies continuously with $\theta$, following the theoretical prediction $P_{10}=\sin^2(\theta/2)$ (green line). The population of $|11\rangle$ ($P_{11}$, red points) also remains expected $P_{11}=\cos^2(\theta/2)$ (red line). Additionally, the insets in Fig.~\ref{fig tunable}(d) show experimental truth tables of Cphase, Controlled Hadamard (CH) and CNOT gate when $\theta$ reaches 0, $\pi/2$, and $\pi$ respectively.
To evaluate the tunability of the phase $\phi$ in the CU operation, we measure the controlled phase after applying a CU operation with $\theta=\pi$ and varied $\phi$. The qubit pair was initially prepared in the state $|1+\rangle$, and a rotation layer ($\rm X^{01}_{\pi/2}$) was applied to project the states for measurements.
As a result, $P_{10}$ exhibits sinusoidal oscillations with a period of $T = \pi$ when $\varphi \in [0,2\pi]$, consistent with theoretical prediction $P_{10}=\cos^2(\phi')$, where $\phi'=\phi+\phi_0$ includes the single-qubit phase $\phi_0$ . Similarly, the population of $|11\rangle$ ($P_{11}$) follows the theoretical prediction $P_{11} = \sin^2(\phi')$.

These results demonstrate that our scheme allows for operation selection within a continuous range by simply adjusting the $\rm X^{12}_{\theta,\varphi}$ gate. The advanced control techniques developed in single-qubit studies ensure high accuracy and reliability of our CU operation. The ability to switch freely simplifies the physical implementation of highly expressive gates.

\subsection{\label{sec:level33} Characterization based on QPT}

\begin{figure}
\centering
\includegraphics[width=1.0\linewidth]{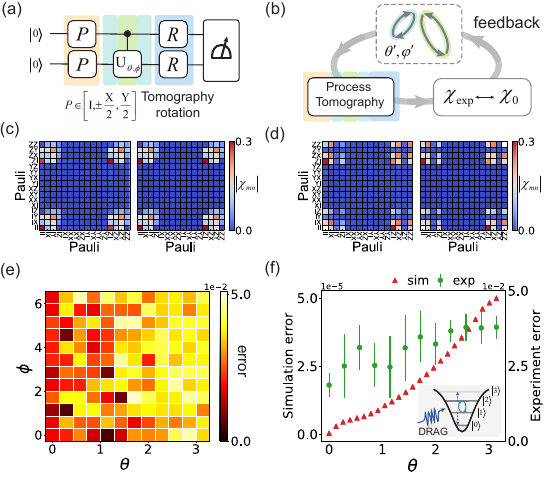}
\caption{\justifying
The characterization of the TCG-CU gate.
(a) Gate sequences for QPT.
(b) The diagram of optimization protocol incorporating QPT and feedback mechanism. 
New parameters are applied to the next pulse based on a comparison between $\chi_0$ and $\chi_{\rm exp}$.
(c) and (d) show the comparison between experimental (left) and theoretical (right) $\chi$ for CH gate and CNOT gate respectively, incorporating single-qubit phases.
(e) Error rates obtained from QPT for the CU family. 
Each data point undergoes 10 QPT experiments.
Error rates remain below 4.8\%, which increase with $\theta$ (0-$\pi$) and have no obvious correlation with $\phi$ (0-2$\pi$).
(f) The slice of (e) results at varying $\theta$.
Simulation (red) and experimental  results (green) were compared, both showing a degradation trend as $\theta$ increases.
The small diagram on the left illustrates leakage induced by DRAG waveforms.
}
\label{fig QPT}
\end{figure}

We optimized the fidelities of the CU family using both the standard quantum process tomography (QPT) and a feedback-optimized protocol, as depicted in Figs.~\ref{fig QPT}(a) and 3(b), respectively.
In this protocol, the experimental quantum process matrices ($\chi_{\rm exp}$) are obtained from QPT and analyzed to extract the operation matrices. Updated $\rm X^{12}$ operations are then applied iteratively until $\chi_{\rm exp}$ converges towards the expected process matrix ($\chi_{0}$).
For instance, figure~\ref{fig QPT}(c) displays the final $\chi_{\rm exp}$ and $\chi_{0}$ for a CNOT gate, where the measured $\chi_{\rm exp}$ exhibits the anticipated block structure and a fidelity of $97.53\pm 0.82\%$ as determined by $F = {\rm Tr}(\chi_{\rm exp} \chi_{0})$. Similarly, Figure~\ref{fig QPT}(d) shows a CH gate achieving a fidelity of $98.07\pm 1.02\%$.

Subsequently,
the fidelities of the CU family were determined using the process described above and are presented in Fig.~\ref{fig QPT}(e).
Experimental results indicate that the fidelities of these operations range from 95.2\% to 99.0\%.
Under identical setups, numerical simulations performed with the QUTIP package yield error rates below $5\times 10^{-5}$~\cite{johansson2012qutip}.
Additionally, we extracted a slice from the fidelity map and compared one representative line (green) with simulation results (red), as shown in Fig.~\ref{fig QPT}(f).
Both the experimental and simulated fidelities gradually decrease as $\theta$ increases, revealing that the primary error source is the imperfect $\rm X^{12}$ operation.
Notably, the conventional DRAG scheme exhibits residual leakage to the states $|0\rangle$ and $|3\rangle$ (see the inset in Fig.~\ref{fig QPT}(f)). Furthermore, imperfections in two-qubit operations and decoherence significantly contribute to the fidelity degradation, as detailed in Appendix~\ref{Error Analysis}.
Finally, these results also encompass errors from initial state preparation and tomography rotations. Improved  control techniques and noise-free environments are expected to mitigate these issues and further broaden the applicability of our scheme.

\subsection{\label{sec:level34} Entangled state prepration}

Entangled states are fundamental to quantum algorithms, quantum communications, and quantum metrology~\cite{horodecki2009quantum,song2019generation,cao2023generation}. However, as the number of qubits increases, the corresponding digital preparation circuits become increasingly  complex, and the cumulative effects of gate errors and decoherence can significantly degrade state fidelity. Employing a more expressive gate set can effectively reduce circuit depth, thereby mitigating these detrimental effects. Compared to conventional CZ-based circuits, our CU-based preparation of three-qubit W and GHZ states achieves circuit depth reductions of 40\% and 44\%, respectively. As the qubit number (\(m\)) increases, these reductions approach approximately 50.0\% and 66.7\%, respectively—a detailed comparison is provided in Table~\ref{tab entanglement}. We next present the preparation circuits for two types of three-qubit entangled states under the TCG scheme, described below:
\begin{gather}
|\Psi_{\rm GHZ}\rangle = \frac{1}{\sqrt{2}} \left( 
|000\rangle + e^{i\tau} |111\rangle
\right), \\
|\Psi_{\rm W}\rangle = \frac{1}{\sqrt{3}} \left( 
|001\rangle + \lambda |010\rangle
+ \sqrt{2-\lambda^2} |001\rangle
\right),
\end{gather}
where $\tau \in \left[0,2\pi\right]$ and $\lambda \in \left[0,\sqrt{2}\right]$ correspond to the GHZ and W state families.

\begin{table}[t]
\centering
\begin{tabular}{cccc}
\toprule[1pt]
\midrule
\textbf{} & \textbf{$N_{1q}$} & $N_{2q}$ & \textbf{$d$} \\ \midrule
GHZ (CZ) & $2m-1$ & $m-1$ & $2m-1$\\  
W (CZ) & $5m-6$ & $3m-5$ & $6m-9$\\ \midrule 
GHZ (CU) & 1 & $m-1$ & $m$\\ 
W (CU) & 2 & $2m-3$ & $2m-1$ \\ \midrule
\bottomrule[1pt]
\end{tabular}
\captionsetup{justification=raggedright,singlelinecheck=false}
\caption{
Comparison of the number of single-qubit gates ($N_{1q}$), two-qubit gates ($N_{2q}$), and circuit depth ($d$) in entangled state preparation circuits with m qubits ($m\geq3$).
We should consider the topology of the QPU for actual state preparation quantum circuits.
For simplicity, we only consider unidirectional preparation methods on the one-dimensional qubit chain.
}
\label{tab entanglement}
\end{table}

\begin{figure}
\centering
\includegraphics[width=1.0\linewidth]{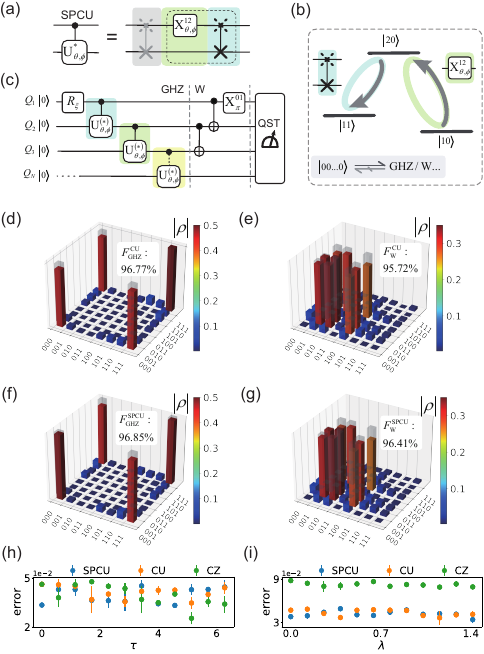}
\caption{\justifying
    (a) Illustration of the composition of SPCU gate under SPTCG scheme.
    The first $\sqrt{\rm CZ}$ gate is absent from the CU operation.
    (b) The state evolution diagram corresponding to (a).
    Unlike the full CU operation, the evolution path of SPCU is unidirectional.
    (c) Quantum circuits for preparing and characterizing GHZ and W states.
    Colored controlled operations in circuits are implemented using TCG or SPTCG schemes, while the others are realized using the TCG scheme.
    $R_{\xi}$ is an ${\rm SU_2}$ single-qubit gate composed of two $\rm X_{\pi/2}^{01}$ gates, treated as a single operation here for consistency.
    (d) and (e) respectively represent experimental $\rho$ of the standard GHZ and W states based on TCG scheme, while (f) and (g) are obtained from SPTCG scheme.
    Theoretical predictions are marked in gray.
    (h) depicts a comparison of $F_{|\Psi_{\rm GHZ}\rangle}$ with different $\tau$ constructed using the SPCU, CU and CZ opertions,
    and (i) shows that of $F_{|\Psi_{\rm W}\rangle}$ with various $\lambda$.
    Each data point undergoes 20 QST experiments.
} 
\label{fig entaglement}
\end{figure}

We employ CU gates as entangling operations to generate  families of GHZ and W states, and use standard quantum state tomography (QST) to benchmark their performance. 
Since state preparation involves a unidirectional transition from one quantum state to another, the SPCU gate can be efficiently applied, as illustrated in Fig.~\ref{fig entaglement}(a). 
Figure~\ref{fig entaglement}(c) shows the corresponding quantum circuits to prepare entanglement states, where the colored operations implement the TCG scheme. The W-preparation circuits require additional controlled and X gates compared to the GHZ-preparation circuits.
We then reconstruct the state density matrix ($\rho$) via QST and calculate the state fidelities using the formula $F=\langle\Psi|\rho|\Psi\rangle$, where $|\Psi\rangle$ is the expected state.
The QST results for the standard GHZ and W states under CU (SPCU) operations are presented in Figs.~\ref{fig entaglement}(d)–(g), yielding fidelities of 96.77\% (96.85\%) and 95.72\% (96.41\%), respectively. Furthermore, the fidelities of the states $|\Psi_{\rm GHZ}\rangle$ and $|\Psi_{\rm W}\rangle$ are benchmarked in Figs.~\ref{fig entaglement}(h)–(i), with values ranging from $95.0\%$ to $97.0\%$. 
These results also indicate that SPCU-based preparation offers a slight performance advantage over CU-based preparation due to its shallower circuit depth. 
In preparing states $|\Psi_{\rm GHZ}\rangle$ and $|\Psi_{\rm W}\rangle$, our scheme achieves average improvements of 1.5\% and 4.2\%, respectively, compared to conventional CZ-based circuits.  This advantage is expected to become increasingly significant for large-scale entangled state preparation and other complex quantum circuits, where circuit depth reduction is further amplified.

\subsection{\label{sec:level35}Quantum comparator}

\begin{figure}
\centering
\includegraphics[width=1.0\linewidth]{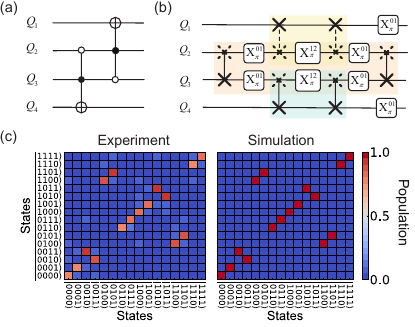}
\caption{\justifying
(a) The equivalent circuit of a single-bit quantum comparator, constructed using two multi-controlled gates.
(b) The detailed implementation of the quantum comparator under the TCG scheme, comprising multiple nested CU gates distinguished by different colors.
(c) Comparison of the experimental and theoretical truth tables for the quantum comparator, with each data point derived from 5000 measurement shots.
} 
\label{fig Qcomparator}
\end{figure}

The quantum comparator, composed of multiple-qubit conditional operations, is the essential component in various quantum algorithms, such as Grover’s algorithm and quantum logic circuits~\cite{oliveira2007quantum,xia2018efficient,xia2019novel}. It essentially functions as a quantum decision-making unit, determining the relationship between two quantum states and adjusting the output accordingly. 
The basic structure of a single-qubit quantum comparator is depicted in Fig.~\ref{fig Qcomparator}. This comparator compares the magnitude of two quantum states \(|a\rangle\) and \(|b\rangle\) in qubits \(Q_1\) and \(Q_2\), outputting the result to qubits \(Q_3\) and \(Q_4\), as summarized in Table~\ref{tab Qcomparator}. However, traditional quantum comparators based on Clifford decompositions are prone to decoherence and gate error accumulation due to their complex and deep circuit structures.

\begin{table}[t]
\centering
\begin{tabular}{ccccc}
    \toprule[1pt]
    \midrule
    \textbf{Function} & $Q_2$ & $Q_3$ & $Q_1$ &  $Q_4$ \\ \midrule
    $a=b$ & $|0\rangle$& $|0\rangle$ & $|0\rangle$ & $|0\rangle$  \\  
    $a=b$ & $|1\rangle$& $|1\rangle$ & $|0\rangle$ & $|0\rangle$  \\ \midrule 
    $a>b$ & $|1\rangle$& $|0\rangle$ & $|1\rangle$ & $|0\rangle$  \\ 
    $a<b$ & $|0\rangle$& $|1\rangle$ & $|0\rangle$ & $|1\rangle$  \\ \midrule
    \bottomrule[1pt]
\end{tabular}
\captionsetup{justification=raggedright,singlelinecheck=false}
\caption{
The function of the quantum comparator is to compare the quantum states of qubits $Q_2$ ($|a\rangle$) and $Q_3$ ($|b\rangle$). Based on this comparison, it either outputs the corresponding states to qubits $Q_3$ and $Q_4$ (their initial states are $|00\rangle$) or applies specific operations on them. 
}
\label{tab Qcomparator}
\end{table}

\begin{table}[t]
\centering
\begin{tabular}{cccc}
    \toprule[1pt]
    \midrule
    \textbf{} & \textbf{$N_{1q}$} & $N_{2q}$ & \textbf{$d$} \\ \midrule
    \textbf{Clifford} & 22 & 12 & 25\\  
    \textbf{TCG} & 8 & 6 & 7\\  \midrule
    \bottomrule[1pt]
\end{tabular}
\captionsetup{justification=raggedright,singlelinecheck=false}
\caption{
Comparison of the number of single-qubit gates ($N_{1q}$), two-qubit gates ($N_{2q}$), and circuit depth ($d$) in quantum comparator construction.
In practical Clifford-based schemes, the actual gate complexity exceeds the values shown in the table due to the additional overhead from decomposing nonlocal controlled gates.
}
\label{tab Qcomparator depth}
\end{table}

To address these issues, we construct and demonstrate a four-qubit single-bit quantum comparator using the TCG scheme. As shown in Fig.~\ref{fig Qcomparator}, the architecture adopts a trapezoidal structure with inherent symmetry, ensuring unitarity. Multiple CU operations are nested to collectively access an auxiliary subspace, before coherently returning to the $\rm S_c$. Two additional single-qubit gates are introduced to facilitate state transitions. Compared to the traditional Clifford decomposition approach, our design achieves significant reductions in circuit complexity (41.18\%) and depth (28.00\%), as detailed in Table~\ref{tab Qcomparator depth}.

Experimentally, we compared the theoretical (\(M_0\)) and experimental (\(M_e\)) results of the comparator truth table (Fig.~\ref{fig Qcomparator}(c)) and observed excellent agreement.
The achieved fidelity was approximately $71.06\%\pm 2.4\%$, calculated using the following formula: 
\begin{align}
F =  \left(\operatorname{Tr} \left( M_e^\dagger M_e \right) + \left| \operatorname{Tr} \left( M_0^\dagger M_e \right) \right|^2 \right) / \left(d (d + 1) \right),
\end{align} 
where $d=16$. In contrast, attempts to construct comparators using CZ gates failed due to noise accumulation and errors in the deeper circuits. These results highlight the robustness and scalability of our scheme, demonstrating its clear advantages and potential in constructing complex quantum circuits.

\section{\label{sec:level4}Discussion and conclusion}
 
In summary, we proposed the TCG scheme and experimentally demonstrated the CU gate and its applications on a superconducting transmon platform, which offers convenient control and reduced complexity costs. Our scheme leverages high-energy levels as auxiliary states instead of auxiliary qubits, enabling independent control freedom via internal operations and transition conversions. These highly expressive and easily manipulated operations serve as key components for significantly reducing quantum circuit depth, thereby mitigating decoherence effects and gate error accumulation. Furthermore, the short-path scheme can further lower the complexity of various quantum circuits, such as state preparation circuits, making it highly suitable for large-scale quantum communications and simulations.

The sandwich-structured CU gate was implemented using the QPT method with a feedback loop, achieving fidelities ranging from 95.2\% to 99.0\%. Detailed slice analysis indicates that residual leakage from high-level single-qubit gates is a primary source of error, while leakage and decoherence in two-qubit gates also contribute to the overall gate error. In three-qubit GHZ and W state preparation, CU-based circuits reduce circuit depth by 40.0\% and 44.0\%, respectively, compared to CZ-based implementations, resulting in fidelity improvements of approximately 1.5\% and 4.2\%. For larger qubit numbers, these reductions approach approximately 50.0\% and 66.7\%. Additionally, we experimentally demonstrated a quantum comparator, achieving a fidelity of approximately 71.2\% with a 72.0\% reduction in circuit depth. These results underscore the substantial potential of our scheme for optimizing quantum circuits, especially in the context of the NISQ era, where minimizing circuit depth is critical.

We also note that the transient occupation of high-energy levels within the non-uniform energy level structure provides additional degrees of freedom for gate construction, enabling the realization of more complex controlled operations. 
However, further exploration of model-based approaches is necessary to move beyond manually constructed designs and achieve arbitrary gate implementations. Quantum Signal Processing presents a promising avenue in this regard.
Meanwhile, increased control errors and decoherence at high-energy levels may degrade the operational fidelity of the TCG scheme. To fully unlock its potential, advanced qubit design strategies and waveform optimization techniques will be essential to mitigate these challenges and enhance performance.

\begin{acknowledgments}

This work has been supported by the National Key Research and Development Program of China (Grant No. 2023YFB4502500), and the National Natural Science Foundation of China (Grant No. 12404564). This work is partially carried out at the USTC Center for Micro and Nanoscale Research and Fabrication. 

\end{acknowledgments}

\appendix
\section{Theoretical Framework}{\label{Theoretical Framework}}

Quantum computing systems often utilize natural atoms (e.g., ion traps) or artificial atoms (e.g., superconducting systems) as experimental platforms ~\cite{koch2007charge,bruzewicz2019trapped,barends2013coherent,ni2023swap,cerfontaine2020high,mooij1999josephson}.
The specific energy levels of these atoms are encoded as qubits for logical computations.
The Hilbert space of $m$ atoms (${\rm S}_{m}$) is defined as the tensor product of the single-qutrit spaces, represented as $|\mathcal{I}\rangle^{\otimes m} (\mathcal{I}\in {\rm N^*}$) . 
In practice, a system of $m$ qutrits can be treated as a single generalized qutrit, allowing ${\rm S}_{m}$ to be partitioned into ${\rm S}_{\rm c}$ and ${\rm S}_{\rm uc}$.
The state vector in ${\rm S}_{\rm c}$ can be expressed as $ |\alpha\rangle = |\mathcal{J}\rangle^{\otimes m} \ (\mathcal{J}\in \{0,1\})$. 
${\rm S}_{\rm uc}$ then forms the complement of the entire space ($\rm C_{S_\mathrm{m}}S_c$), with its state vector described as $|\beta\rangle $. Consequently, the state in the total space is represented as $ |\Psi\rangle = a|\alpha\rangle + b|\beta\rangle$ $(a^2+b^2=1)$.

\begin{figure}[ht]
\centering
\includegraphics[width=0.8\linewidth]{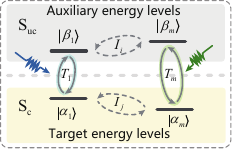}
\caption{\justifying
    The principle diagram of the TCG scheme.
}
\label{fig S principle}
\end{figure}

Our implementation constructs a sequence that comprises transition operations ($T$) between $\rm S_c$ and $\rm S_{uc}$ and internal operations ($I$)  that act within either subspace, as shown in Fig.~\ref{fig S principle}. First, owing to distinct operational conditions and the involvement of different energy levels, multiple independent conversion operations are available for transition between $\rm S_{c}$ and $\rm S_{uc}$, such as an $\rm X^{12}$ gate between states $|1\rangle$ and $|2\rangle$ or an $\mathrm{iSWAP}_{20\leftrightarrow 11}$ gate between states $|20\rangle$ and $|11\rangle$ (also named $\sqrt{\rm CZ}$ gate)~\cite{wang2021optimal,chen2023transmon,chu2023scalable}. Second, internal operations that modify the qubits state within $\rm S_{c}$ or $\rm S_{uc}$ are inherently independent due to the non-uniform energy level structures and Bohr frequency conditions, with Pauli gates serving as typical examples.

Due to specific physical mechanisms, $T$ operations are diverse and exhibit state selectivity.
For instance, $\rm X^{12} $ operation facilitates the transition between states $|1\rangle$ and $|2\rangle$ in a single qutrit, 
while $\sqrt{\rm CZ}$ gate swaps the states $ |11\rangle $ and $|20\rangle $ ($|02\rangle $) in a qutrit pair, 
while leaving other states unchanged.
The process can be described as follows:
\begin{gather}
T_i(|\alpha_i\rangle+|\alpha_j\rangle)=
T_i|\alpha_i\rangle+I_0|\alpha_j\rangle=|\beta_i\rangle+|\alpha_j\rangle,
\end{gather}
where $I_0$ is the identity operation.

In addition,  $I$ operations have distinct impacts within ${\rm S}_{\rm c}$ and ${\rm S}_{\rm uc}$ due to the non-uniform level structure of the qutrit.
For example, $\rm X^{01}$ gate does not induce transitions in the qutrit that is in $ |2\rangle $  or higher state, 
while a $\rm Z$ gate imparts different phases to qutrits depending on their specific states.
The process can be decribed as follows:
\begin{gather}
I_i(|\alpha_i\rangle+|\beta_i\rangle)=
I_i|\alpha_i\rangle+I_0|\beta_i\rangle=|\alpha_j\rangle+|\beta_i\rangle,
\end{gather}
or
\begin{gather}
I'_i(|\alpha_i\rangle+|\beta_i\rangle)=
I_0|\alpha_i\rangle+I'_i|\beta_i\rangle=|\alpha_i\rangle+|\beta_j\rangle.
\end{gather}

Quantum operations rely on establishing connections between target energy levels to achieve controlled energy exchange or phase accumulation.
Utilizing the high-dimensional Hilbert space enables more localized effects of certain operations. Consequently, the combination of these two types of operations achieves a high degree of freedom in constructing conditional operations.
Transition pathways describe the map of combining operations to connect energy levels. By engineering transition pathways, we can achieve highly expressive controlled operations, which is named the Transition Composite Gate (TCG) scheme.

The TCG scheme connects target energy levels $|\alpha_0 \rangle$ and $|\alpha_m\rangle$ through intermediate levels $|\beta_i\rangle$, which typically reside in ${\rm S}_{\rm uc}$. 
Due to the isolation between different subspaces, 
there are mutually independent operations in ${\rm S}_{\rm uc}$ and ${\rm S}_{\rm c}$.
We first send qubits into ${\rm S}_{\rm uc}$ for localization, 
and then establish bridges in ${\rm S}_{\rm uc}$ or ${\rm S}_{\rm c}$.
The above asymmetric processes ($\rm U_{\rm as}$) can be described as follows:
\begin{gather}
{\rm U}_{\rm as} = \prod_{i=1}^{N-1}(T_iI_i)\cdot T_N,
\end{gather}

Since the transition operations do not commute ($T_iT_j \neq T_jT_i,\ i\neq j$), $U_{\rm as}$ is non-unitary:
\begin{gather}
\prod_{i=1}^{N-1}(T_iI_i)\cdot T_N \neq 	
T_N\cdot \prod_{j=1}^{N-1}(I_{N-j}T_{N-j})
\end{gather}
This results in a unidirectional transition from states $|\alpha_0\rangle$ to $|\alpha_m\rangle$ in ${\rm S}_{\rm c}$, 
while state $|\alpha_m\rangle$ transitions to ${\rm S}_{\rm uc}$ due to $T_N$.
Considering the requirement for unitarity of quantum operations, we symmetrize transition pathways ($\rm U_{\rm s}$):
\begin{gather}
{\rm U}_{\rm s}= \prod_{i=1}^{N-1}(T_iI_i)\cdot T_N\cdot
\prod_{j=1}^{N-1}(I_{N-j}T_{N-j}),
\end{gather}
which comprises a total of $2N-1$ $T$ operations.	$I$ operations are not essential and only become relevant after $T$ operations have established spatial isolation. These operations link target energy levels in ${\rm S}_{\rm c}$ by serially utilizing energy levels in ${\rm S}_{\rm uc}$.
Under this symmetrical construction sequenqce, $\rm U_{\rm s}$ satisfies that $\rm  U_{\rm s}U_{\rm s}^{\dag}$ is unitary in ${\rm S}_{\rm c}$.
Here, $\rm  U_{\rm s}$ and $\rm  U_{\rm s}^{\dag}$ correspond to independent transition processes with initial states $|\alpha_0\rangle$ and $|\alpha_m\rangle$, respectively.
Meanwhile, different $T_i$ and $I_i$ can provide $\rm U_{\rm s}$ with degrees of freedom as independent channels.

However, unidirectional evolution processes have been demonstrated to facilitate state preparation and other quantum circuits, thereby further reducing circuit depth. In light of these considerations, we propose an asymmetric short-path TCG (SPTCG) scheme formulated as:
\begin{gather}
{\rm U}_{\rm {SP}}= \prod_{i=1}^{N-1}(T_iI_i)\cdot T_N.
\end{gather}
Compared to the full scheme, target energy levels are connected unidirectionally here (from $|\alpha_0\rangle$ to $|\alpha_m\rangle$). As a result, the SPTCG scheme reduces the number of operations by almost half without losing functionality. 
For example, in state preparation, SPTCG can further shallow the circuit without losing functionality.

Moreover, transition operations are not restricted to a limited dimension. This implies that the TCG scheme has the potential to implement qutrit and multi-qubit operations. We provide some examples in the next section.

\section{Examples}{\label{Examples}}

We employ several quantum operations as examples.
These operations are represented in the space spanned by the states $|00\rangle$, $|01\rangle$, $|02\rangle$, $|10\rangle$, $|11\rangle$, and $|12\rangle$, which are described as Eqs.~\eqref{CP}, ~\eqref{X01}, and ~\eqref{X12}.

\begin{figure*}
\begin{align}
{\rm CP}_{\theta}&=
\begin{pmatrix}
    1 & 0 & 0 & 0 & 0 & 0	\\
    0 & 1 & 0 & 0 & 0 & 0	\\
    0 & 0 & \cos(\frac{\theta}{2}) & 0 & -i\sin(\frac{\theta}{2}) & 0	\\
    0 & 0 & 0 & 1 & 0 & 0			\\
    0 & 0 & -i\sin(\frac{\theta}{2}) & 0 & \cos(\frac{\theta}{2}) & 0 \\
    0 & 0 & 0 & 0 & 0 & 1	
\end{pmatrix}, \label{CP}\\
I_0 \otimes {\rm X}_{\theta',\phi'}^{01}&=
\begin{pmatrix}
    \cos(\frac{\theta'}{2}) & e^{-i\phi'} \sin(\frac{\theta'}{2}) & 0 & 0 & 0 & 0 	\\
    e^{i\phi'} \sin(\frac{\theta'}{2})  & \cos(\frac{\theta'}{2}) & 0 & 0 & 0 & 0	\\
    0 & 0 & 1 & 0 & 0 & 0 \\
    0 & 0 & 0 & \cos(\frac{\theta'}{2}) & e^{-i\phi'} \sin(\frac{\theta'}{2})  & 0  \\
    0 & 0 & 0 & e^{i\phi'} \sin(\frac{\theta'}{2})  & \cos(\frac{\theta'}{2}) & 0  \\
    0 & 0 & 0 & 0 & 0 & 1
\end{pmatrix},\label{X01}\\
I_0 \otimes {\rm X}_{\theta'', \phi''}^{12}&=
\begin{pmatrix}
    1 & 0 & 0 & 0 & 0 & 0 	\\
    0 & \cos(\frac{\theta''}{2}) & e^{-i\phi''} \sin(\frac{\theta''}{2}) & 0 & 0 & 0	\\
    0 & e^{i\phi''} \sin(\frac{\theta''}{2}) & \cos(\frac{\theta''}{2}) & 0 & 0 & 0\\
    0 & 0 & 0 & 1 & 0 & 0			\\
    0 & 0 & 0 & 0 & \cos(\frac{\theta''}{2}) & e^{-i\phi''} \sin(\frac{\theta''}{2}) \\
    0 & 0 & 0 & 0 & e^{i\phi''} \sin(\frac{\theta''}{2}) & \cos(\frac{\theta''}{2})
\end{pmatrix},\label{X12}
\end{align}
\end{figure*} where $\theta$ and $\phi$ denote the Rabi angle and phase introduced by the operation. 
CP operation refers to the interaction of qubits at the resonance point between the states $|11\rangle$ and $|02\rangle$.
Additionally, ${\rm CP_{\pi}}={\rm CZ}$, ${\rm CP_{\pi/2}}={\rm \sqrt{CZ}}$.

\begin{figure*}[ht]
\centering
\includegraphics[width=0.75\linewidth]{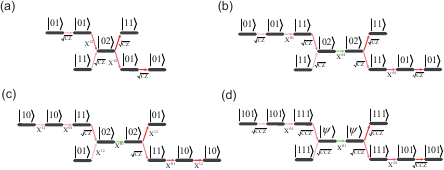}
\caption{\justifying
    Illustration of the transition paths of several TCG-based schemes.
    (a) and (b) show the transition pathways for two types of CU gate operations. 
    (c) and (d) correspond to pathways for the SWAP family and CCU gate, respectively.
}
\label{fig S paths}
\end{figure*}

\textbf{i) CU family.}
First, we derive the TCG-based CU operation as detailed in the main text.
Its transition paths are shown in Fig.~\ref{fig S paths}(a).
Its mathematical description in the full two-qubit Hilbert space ($|00\rangle$, $|01\rangle$, $|02\rangle$, $|10\rangle$, $|11\rangle$, $|12\rangle$, $|20\rangle$, $|21\rangle$ and $|22\rangle$) is given by  Eq.~\eqref{CU_com}.

\begin{figure*}
\begin{gather}
\rm {CU}= \sqrt{\rm CZ} \cdot \mathit{I}_0 \otimes X_{\theta, \phi}^{12} \cdot \sqrt{\rm CZ}\\
= 
\begin{pmatrix}
    1 & 0 & 0 & 0 & 0 & 0 & 0 & 0 & 0 	\\
    0 & \cos(\frac{\theta}{2}) & 0 & 0 & -ie^{-i\phi}\sin(\frac{\theta}{2}) & 0 & 0 & 0 & 0 	\\
    0 & 0 & -\cos(\frac{\theta}{2}) & 0 & 0 & -ie^{-i\phi}\sin(\frac{\theta}{2}) & 0 & 0  & 0 	\\
    0 & 0 & 0 & 1 & 0 & 0 & 0 & 0 & 0 	\\
    0 & -ie^{i\phi}\sin(\frac{\theta}{2}) & 0 & 0 & -\cos(\frac{\theta}{2}) & 0 & 0 & 0 & 0 	\\
    0 & 0 & -ie^{i\phi}\sin(\frac{\theta}{2}) & 0 & 0 & \cos(\frac{\theta}{2}) & 0 & 0 & 0 	\\
    0 & 0 & 0 & 0 & 0 & 0 & 1 & 0 & 0 	\\
    0 & 0 & 0 & 0 & 0 & 0 & 0 & \cos(\frac{\theta}{2}) & e^{-i\phi}\sin(\frac{\theta}{2})	\\
    0 & 0 & 0 & 0 & 0 & 0 & 0 & e^{i\phi}\sin(\frac{\theta}{2}) & \cos(\frac{\theta}{2}) 	\\
\end{pmatrix}.\label{CU_com}
\end{gather}
\end{figure*}

For qubits, the computational states are within ${\rm S}_{\rm c}$ spanned by the states $|0\rangle$ and $|1\rangle$.
Therefore, the above matrix can be reduced to the subspace ${\rm S}_{\rm c}$ ($|00\rangle$, $|01\rangle$, $|10\rangle$ and $|11\rangle$):
\begin{gather}
\rm {CU}
= 
\begin{pmatrix}
    1 & 0 & 0 & 0 	\\
    0 & \cos(\frac{\theta}{2}) & 0 & -ie^{-i\phi}\sin(\frac{\theta}{2})	\\
    0 & 0 & 1 & 0 	\\
    0 & -ie^{i\phi}\sin(\frac{\theta}{2}) & 0 & -\cos(\frac{\theta}{2})
\end{pmatrix}.
\end{gather}
In this CU gate, $\theta$ and $\phi$ are directly associated with the parameters of $\rm X_{\theta, \phi}^{12}$, 
which can be precisely controlled using well-established single-qubit control techniques.
In fact, we can endow this operation with additional degrees of freedom.
We can apply SNZ schemes with different parameters to add a conditional phase $\phi_{c1}$ to $|11\rangle\langle 11|$~\cite{rol2019fast}.
Additionally, we can introduce an extra Z gate with phase $\phi_{c2}$,
that is, 
by replacing $\rm I_0\otimes X^{12} $ with $\rm Z(\phi_{c2})\otimes Z\phi_{c1}\cdot X^{12} $, 
thereby adding an additional controlled phase $\phi_{c2}$ to $|01\rangle\langle 01|$.
Hence, the final matrix can be described as:
\begin{align}
\rm {CU} &= \rm \sqrt{CZ} \cdot Z(\phi_{c2})\otimes Z(\phi_{c1})X^{12} \cdot \sqrt{CZ} \\
&= 
\begin{pmatrix}
    1 & 0 & 0 & 0 	\\
    0 & \cos(\frac{\theta}{2}) & 0 & -ie^{-i\phi}\sin(\frac{\theta}{2})	\\
    0 & 0 & e^{i\phi_{c2}} & 0 	\\
    0 & -ie^{i\phi}\sin(\frac{\theta}{2}) & 0 & -e^{i\phi_{c1}}\cos(\frac{\theta}{2})
\end{pmatrix}.
\end{align}

\textbf{ii) Another CU family implementation.}
We use an alternative combination to implement the CU gate to illustrate the TCG scheme of weaving operations in ${\rm S}_{\rm c}$.
Figure.~\ref{fig S paths}(b) illustrates the transfer process,
which can be described as Eq.~\eqref{CU_com2}.
\begin{figure*}
\begin{gather}
\rm {CU'}= \sqrt{\rm CZ}_{\pi} \cdot \mathit{I}_0 \otimes X_{\pi,\phi_1}^{01} \cdot {\rm CP}_{\theta} \cdot \mathit{I}_0 \otimes X_{\pi,\phi_2}^{01} \cdot \sqrt{\rm CZ}_{\pi}\\
= 
\begin{pmatrix}
    1 & 0 & 0 & 0 & 0 & 0 & 0 & 0 & 0	\\
    0 & e^{2i(\phi_1-\phi_2)} & 0 & 0 & 0 & 0 & 0 & 0 & 0	\\
    0 & 0 & -e^{2i(\phi_1-\phi_2)} & 0 & 0 & 0 & 0 & 0 & 0	\\
    0 & 0 & 0 & \cos(\frac{\theta}{2}) &  -e^{-i\phi_2}\sin(\frac{\theta}{2}) & 0 & 0 & 0 & 0	\\
    0 & 0 & 0 & -e^{i\phi_1}\sin(\frac{\theta}{2}) & -e^{i(\phi_1-\phi_2)}\cos(\frac{\theta}{2}) & 0 & 0 & 0 & 0	\\
    0 & 0 & 0 & 0 & 0 & e^{i(\phi_1-\phi_2)} & 0 & 0 & 0	\\
    0 & 0 & 0 & 0 & 0 & 0 & 1 & 0 & 0	\\
    0 & 0 & 0 & 0 & 0 & 0 & 0 & e^{2i(\phi_1-\phi_2)} & 0	\\
    0 & 0 & 0 & 0 & 0 & 0 & 0 & 0 & e^{i(\phi_1-\phi_2)}\\
\end{pmatrix}.\label{CU_com2}
\end{gather}
\end{figure*}

It can be observed that the portion of the matrix corresponding to ${\rm S}_{\rm uc}$ remains diagonal because the operations are performed within ${\rm S}_{\rm c}$.
The $\rm X^{01}$ operation affects only ${\rm S}_{\rm c}$, 
while ${\rm S}_{\rm uc}$ remains unaffected due to the isolation between ${\rm S}_{\rm c}$ and ${\rm S}_{\rm uc}$. 
However, since the restoration of state $|00\rangle$ and state $|01\rangle$ in ${\rm S}_{\rm c}$ must be considered,
the number of $\rm X^{01}$ operations is even.
This CU operation can also be restricted within ${\rm S}_{\rm c}$:
\begin{gather}
\rm {CU'}
= 
\begin{pmatrix}
    1 & 0 & 0 & 0 	\\
    0 & e^{2i(\phi_1-\phi_2)} & 0 & 0 	\\
    0 & 0 & \cos(\frac{\theta}{2}) & -e^{-i\phi_2}\sin(\frac{\theta}{2})	\\
    0 & 0 & -e^{i\phi_1}\sin(\frac{\theta}{2}) & -e^{i(\phi_1-\phi_2)}\cos(\frac{\theta}{2})	\\
\end{pmatrix}.
\end{gather}

We also find that this path weaving method can be directly applied to a qutrit system. 
For example, by replacing $\rm X^{01}$ in Eq.~\eqref{CU_com2} with $\rm X^{12}$, we can achieve a controlled exchange between states $|01\rangle$ and $|12\rangle$.
The process is as Eq.~\eqref{CU_qutrit}.

\begin{figure*}
\begin{gather}
\rm {CU}_{\rm qutrit}= \sqrt{\rm CZ}_{\pi}\cdot  \mathit{I}_0 \otimes X_{\pi,\phi_1}^{12} \cdot {\rm CP}_{\theta,\phi_q'}\cdot \mathit{I}_0 \otimes X_{\pi,\phi_2}^{12} \cdot \sqrt{\rm CZ}_{\pi}\\
= 
\begin{pmatrix}
    1 & 0 & 0 & 0 & 0 & 0 & 0 & 0 & 0	\\
    0 & e^{i(-\phi_1+\phi_2)}\cos(\frac{\theta}{2}) & 0 & 0 & 0 & -ie^{-i(\phi_1+\phi_2)}\sin(\frac{\theta}{2}) & 0 & 0 & 0	\\
    0 & 0 & -e^{i(-\phi_1+\phi_2)} & 0 & 0 & 0 & 0 & 0 & 0	\\
    0 & 0 & 0 & 1 & 0 & 0 & 0 & 0 & 0	\\
    0 & 0 & 0 & 0 &  -e^{i(\phi_1-\phi_2)} & 0 & 0 & 0 & 0	\\
    0 & -ie^{i(\phi_1+\phi_2)}\sin(\frac{\theta}{2}) & 0 & 0 & 0 & e^{i(\phi_1-\phi_2)}\cos(\frac{\theta}{2}) & 0 & 0 & 0	\\
    0 & 0 & 0 & 0 & 0 & 0 & 1 & 0 & 0	\\
    0 & 0 & 0 & 0 & 0 & 0 & 0 & e^{i(-\phi_1+\phi_2)} & 0	\\
    0 & 0 & 0 & 0 & 0 & 0 & 0 & 0 & e^{i(\phi_1-\phi_2)}	\\
\end{pmatrix}.\label{CU_qutrit}
\end{gather}
\end{figure*}
Compared to the path design in i), 
the path here involves more components but has less impact on potential leakage propagation. Furthermore, in higher-dimensional qutrits, this scheme can leverage existing components to more independently select any two energy levels for controlled energy exchange.

\textbf{iii) SWAP family.}
We can use the TCG scheme to construct the SWAP gate family, which essentially involves controlled exchanges between the states $|01\rangle$ and $|10\rangle$.
This is an example of simultaneous path design for ${\rm S}_{\rm c}$ and ${\rm S}_{\rm uc}$, composed of $\rm X^{01}$, $\rm X^{12}$, and $\rm CP$ operations, as described in  Eq.~\eqref{SWAP_com}.
\begin{figure*}
\begin{gather}
\rm {SWAP}= \mathit{I}_0 \otimes X_{\pi,\phi_1}^{12} \cdot \mathit{I}_0 \otimes X_{\pi,\phi_2}^{01} \cdot {\rm CP}_{\theta,\phi_q} \cdot \mathit{I}_0 \otimes X_{\pi,\phi_3}^{01} \cdot \mathit{I}_0 \otimes X_{\pi,\phi_4}^{12}\\
= 
\begin{pmatrix}
    1 & 0 & 0 & 0 & 0 & 0 & 0 & 0 & 0	\\
    0 & e^{i\phi_x}\cos(\frac{\theta}{2}) & 0 & -ie^{i(-\phi_1+\phi_2)}\sin(\frac{\theta}{2}) & 0 & 0 & 0 & 0 & 0	\\
    0 & 0 & e^{i\phi_y} & 0 & 0 & 0 & 0 & 0 & 0	\\
    0 & -ie^{i(-\phi_3+\phi_4)}\sin(\frac{\theta}{2}) & 0 & \cos(\frac{\theta}{2}) & 0 & 0 & 0 & 0 & 0	\\
    0 & 0 & 0 & 0 & e^{i\phi_x} & 0 & 0 & 0 & 0	\\
    0 & 0 & 0 & 0 & 0 & e^{i\phi_y} & 0 & 0 & 0	\\
    0 & 0 & 0 & 0 & 0 & 0 & 1 & 0 & 0	\\
    0 & 0 & 0 & 0 & 0 & 0 & 0 & e^{i\phi_x} & 0	\\
    0 & 0 & 0 & 0 & 0 & 0 & 0 & 0 & e^{i\phi_y}	\\
\end{pmatrix}, \label{SWAP_com}
\end{gather}
\end{figure*}
where $\phi_x = -\phi_1+\phi_2-\phi_3+\phi_4$ and $\phi_y = \phi_1+2\phi_2-2\phi_3-\phi_4$.
In ${\rm S}_{\rm c}$,
the operation is Eq.~\eqref{SWAP}.
\begin{figure*}
\begin{gather}
\rm {SWAP}
= 
\begin{pmatrix}
    1 & 0 & 0 & 0 	\\
    0 & e^{i\phi_x}\cos(\frac{\theta}{2}) & -ie^{i(-\phi_1+\phi_2)}\sin(\frac{\theta}{2}) & 0 	\\
    0 & -ie^{i(-\phi_3+\phi_4)}\sin(\frac{\theta}{2}) & \cos(\frac{\theta}{2}) & 0 	\\
    0 & 0 & 0 & e^{i\phi_x} 	\\
\end{pmatrix}, \label{SWAP}
\end{gather}
\end{figure*}

Figure~\ref{fig S paths}(c) illustrates the transition paths where states $|01\rangle$ and $|10\rangle$ sequentially transition to ${\rm S}_{\rm c}$ and return to complete the swap, while all other states remain isolated.
The TCG-based SWAP differs from the resonance-based SWAP as it relies on digital implementation.
Additionally, on our experimental platform, smaller qubit frequency adjustments imply reduced likelihood of energy level collisions in complex environments and minimize Z crosstalk effects.
Meanwhile, it is shallower than the traditional SWAP circuit composed of three CNOT gates, while also possessing higher expressivity.

We also note that the application of the TCG scheme in i), ii), and iii) can unify two-qubit working points (interaction frequencies).
Fewer parameters reduce the initialization time cost of quantum processing unit (QPU) and influence of frequency collisions, 
thereby facilitating the development of highly expressive digital quantum circuits.

\textbf{iv) Multiple-qubit controlled operation.}

\begin{figure}[ht]
\centering
\includegraphics[width=1.0\linewidth]{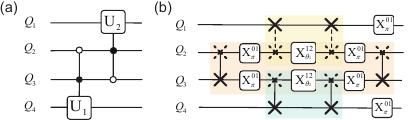}
\caption{\justifying
    An example of a multi-qubit controlled operation where two single-qubit gates ($\rm X^{12}_{\theta_1,\theta_2}$) control the corresponding controlled unitaries ($U_{1,2}$).
}
\label{fig S MCU}
\end{figure}

The implementation of multi-qubit controlled gates can be achieved via either two-qubit interactions or direct multi-qubit interactions.
The first approach leverages two-qubit interactions to induce transition conversions within a high-energy subspace involving multiple qubits. A notable example is the quantum comparator discussed in the main text, which actually exhibits higher expressiveness. The controlled gate is inherently unitary, and the quantum comparator can be regarded as a special case of the $\rm X^{12}_{\pi}$ gate, as illustrated in Fig.~\ref{fig S MCU}.

In contrast, the multi-qubit interaction approach reduces circuit depth while increasing the complexity of the gate components. 
To demonstrate the scalability of the TCG scheme, we present an example of multi-qubit operations in Fig.~\ref{fig S paths}(d).
In Ref.~\cite{li2019realisation}, it is proposed to implement a diabatic CCZ gate by simultaneously resonating state $|111\rangle$ with the states $|021\rangle$ and $|030\rangle$ . 
The population of state $|111\rangle$ is almost temporarily transferred to a superposition state $|\psi\rangle = a|021\rangle+b|030\rangle$ (CCP operation), 
where $a$ and $b$ are complex numbers satisfying $a^2+b^2=1$.
Since $|\psi\rangle$ remains in ${\rm S}_{\rm uc}$, the combined design in ii) can directly implement the CCU gate, as follows:
\begin{gather}
\rm {CCU}= \sqrt{\rm CCZ} \cdot \mathit{I}_0 \otimes X_{\pi,\phi_1}^{01} \cdot {\rm CCP}_{\theta} \cdot \mathit{I}_0 \otimes X_{\pi,\phi_2}^{01} \cdot \sqrt{\rm CCZ}.
\end{gather}
Its transition path is consistent with ii), thereby enabling controlled exchange between states $|101\rangle$ and $|111\rangle$.

\section{Device}{\label{Device}}

\begin{figure*}[ht]
\centering
\includegraphics[width=0.8\linewidth]{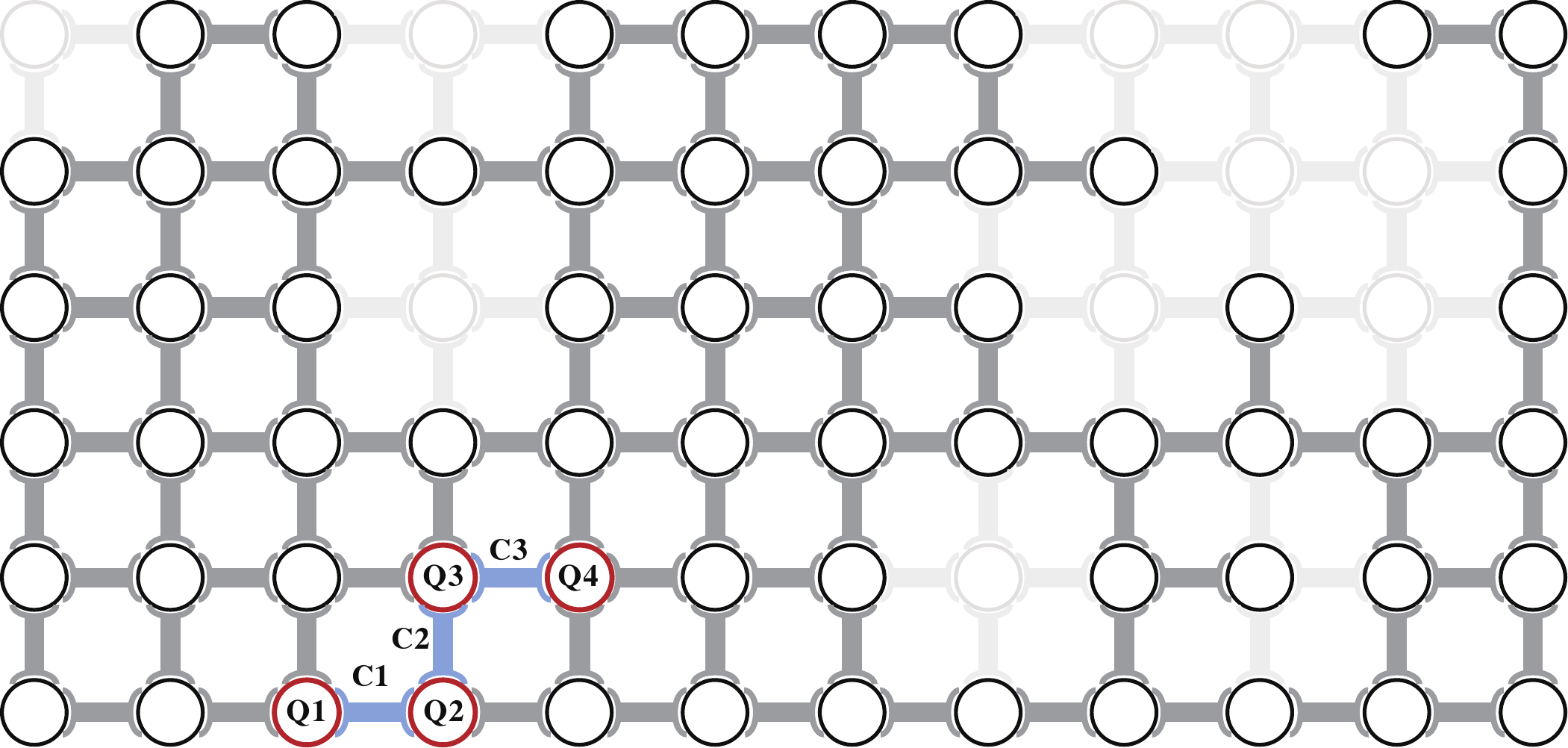}
\caption{\justifying
The topology of the superconducting processor \textit{Wukong} and qubits used in the experiment.
Dark regions indicate well-functioning qubits and couplers, while light regions represent those that are malfunctioning. 
We selected four adjacent qubits (highlighted in red) as our experimental platform.
}
\label{fig S device}
\end{figure*}

We conducted experiments on a superconducting quantum processor named \textit{Wukong}, which comprises 72 transmon superconducting qubits and 126 transmon-type couplers arranged in a grid-like topology.
The topology of the chip is illustrated in Fig.~\ref{fig S device}.
We selected four qubits from this processor as our experimental platform,
and their basic parameters and decoherence performance listed in Table~\ref{tab device}.

\begin{table}[ht]
\centering
\begin{tabular}{ccccc}
    \toprule[1pt]
    \midrule
    \textbf{Parameters} & \textbf{Q1} & \textbf{Q2} & \textbf{Q3} & \textbf{Q4}\\ \midrule
    \textbf{$f_{01}\ \rm (GHz)$} & 4.659 & 4.280 & 4.098 & 3.954\\ 
    \textbf{$f_{12}\ \rm (GHz)$} & 4.421 & 4.030 & 3.852 & 3.710\\ 
    \textbf{$\widetilde{f} \rm (GHz)$} & 4.498/4.055 & 4.324/4.280 & 4.093 & 3.865\\ \midrule 
    \textbf{$T_1\ \rm (\mu s)$} & 11.019 & 10.677 & 11.259 & 13.512\\ 
    \textbf{$\widetilde{T}_1\ \rm (\mu s)$} & 8.604/7.956 & 11.777/10.677 & 11.259 & 13.512\\
    \textbf{$T^*_2\ \rm (\mu s)$} & 6.051 & 2.264 & 3.134 & 2.256 \\ 
    \textbf{$\widetilde{T}^*_2\ \rm (\mu s)$} & 1.403/1.222 & 2.533/2.264 & 3.134 & 2.256\\ \midrule
    \bottomrule[1pt]
\end{tabular}
\captionsetup{justification=raggedright,singlelinecheck=false}
\caption{
    Relevant qubit parameters and performance metrics for the device under test.
    We list qubit transition frequency from $|i\rangle$ to $|j\rangle$ ($f_{ij}$),
    the two-qubit interaction frequency ($\widetilde{f}$),
    the qubit coherence times ($T_1$, $T^*_2$) at the idle point,
    and the qubit coherence times at the two-qubit working point ($\widetilde{T}_1$, $\widetilde{T}^*_2$).
    $Q_3$ and $Q_4$ are placed at the sweetspot, 
    while $Q_1$ and $Q_2$ are positioned at the waist of their magnetic flux frequency modulation curve.
}
\label{tab device}
\end{table}

\section{Error Analysis}{\label{Error Analysis}}

In this section, we perform simulations to investigate the impact of single-qubit leakage and decoherence on TCG-CU operations in the main text.

\begin{figure}[ht]
\centering
\includegraphics[width=1.0\linewidth]{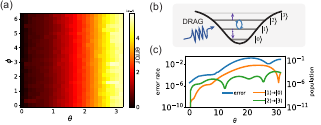}
\caption{\justifying
    (a) The simulation error rates for the CU family. 
    (b) The diagram illustrates leakage induced by DRAG waveforms.
    (c) Simulated data plot illustrating the relationship between error rate and leakage over the range $\theta \in [0, 10\pi]$.
}
\label{fig S simulation QPT}
\end{figure}

\textbf{i) Leakage.} 
The component $\rm X^{12}$ of the CU gate employs the classical Derivative Reduction by Adiabatic Gate (DRAG) waveform scheme to minimize leakage effects~\cite{motzoi2009simple}.
Some studies suggest that using a simple parameterized waveform for $\rm X^{12}$ is insufficient to suppress leakage across multiple energy levels, resulting in transitions from state $|1\rangle$ to state $|0\rangle$ and from state $|2\rangle$ to state $|3\rangle$ (see Fig.~\ref{fig S simulation QPT}(b))~\cite{gambetta2011analytic}.

We simulated the impact of leakage using the QuTiP package with parameters matching those of our experimental platform.
We collected fidelity data and performed slicing analysis, as shown in Fig.~\ref{fig S simulation QPT}(a).
A clear trend was observed showing that fidelity decreases as $\theta$ increases, while exhibiting minimal dependence on the value of $\phi$. This observation is consistent with experimental results.
We also observed that as $\theta$ increased, 
the leakage from state $|1\rangle$ to state $|0\rangle$ and from state $|2\rangle$ to state $|3\rangle$ on $Q_2$ increased continuously over the first three cycles.
Fig.~\ref{fig S simulation QPT}(c) illustrates leakage states for different $\rm X^{12}$ transitions and compares these with the fidelities.
These trends confirm that leakage significantly impacts the fidelity of the CU gate. Extending the DRAG scheme and optimizing qubit frequencies and anharmonicities are potential solutions~\cite{motzoi2009simple,gambetta2011analytic}.

\textbf{ii) Decoherence.} 
Decoherence remains a persistent challenge in quantum computing. In the TCG-CU scheme, qubits are temporarily excited to the state $|2\rangle$, which exhibits faster decoherence relative to $|1\rangle$, specifically $T_1^{|2\rangle} = T_1^{|1\rangle}/\kappa$, where $\kappa$ is a ratio and is usually greater than 1 (we set $\kappa=\sqrt{2}$ here )~\cite{morvan2021qutrit}.
Therefore, compared to the traditional CZ operation, decoherence may have a more serious impact in the TCG scheme.
We conduct simulations to observe the influence of varying degrees of decoherence on TCG-based CU gates and compared it with CZ-based CNOT gates.

\begin{figure}[ht]
\centering
\includegraphics[width=1.0\linewidth]{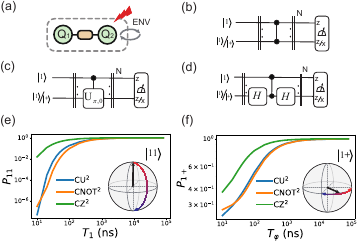}
\caption{\justifying
    (a) The QPU is not a strictly isolated system. 
    Energy exchange with the environment and the impact of noise lead to decoherence.
    (b), (c), and (d) respectively show the circuit diagrams used to test the CZ, CU, and CNOT operations.
    In these experiments, the duration of single-qubit gates are 30 ns, the CZ gate are 40 ns, and the CU gate are 90 ns.
    (e) and (f) respectively show the data graphs depicting the impact of different $T_1$ and $T_\phi$ on the quantum operations.
}
\label{fig S path independent}
\end{figure}

To separately observe effects of longitudinal relaxation and transverse relaxation, we first prepared  qubits in states \(|11\rangle\) and \(|1+\rangle\), applied different gate sequences forming an identity operation, and subsequently measured the states $|11\rangle$ and $|1+\rangle$.
The circuit diagrams and data plots are shown in Fig.~\ref{fig S path independent}.
We observed that the CZ gate exhibited the best performance in decoherence tests, while the CU gate slightly outperformed the CZ-based CNOT gate, primarily due to different duration. 
Considering the high expressiveness demonstrated by the CU gate, it remains advantageous over the traditional CNOT gate in certain quantum circuits.

Improvements in manufacturing processes and upgrades to control systems will continuously mitigate the effects of dissipative coupling and noise factors, thereby enhancing the decoherence performance of qubits.

\section{Path-Independent Properties}{\label{Path-Independent Properties}}

In the TCG scheme, 
transition operations can be optimized for robustness through waveform design, enabling broader applications in the complex environment.

First, 
we note that the traditional diabatic CZ scheme relies on controlled phase accumulation of conditional phase to rapidly establish entanglement.
The simultaneous need to suppress leakage while maintaining controlled phase calls for careful design of evolution paths and considerations of factors such as ZZ crosstalk and other potential sources of error.
Noise, waveform distortions, electromagnetic crosstalk, and observer effects can all cause deviations of the state vector from its expected trajectory on the Bloch sphere, resulting in leakage or insufficient conditional phase~\cite{eroms2006low,julin2010reduction,yuge2011measurement,bylander2011noise,wang2022control,kanaar2024robust,krinner2020benchmarking}.
In large-scale quantum computing systems, the abundance of interference sources prompts the demand for quantum operations that are resilient to a wide array of disturbances.
Recent researches have found that free waveform design contributes to achieving high robustness in quantum operations.
However, dual cost functions of leakage and controlled phase still result in complicated models and a large number of parameters, 
leading to high initialization time costs and complex classical crosstalk corrections .

We find that $\phi_q$ introduced by the two-qubit interaction in the TCG-CNOT operation matrix can be regarded as a correctable single-qubit phase,
specifically:
\begin{gather}
\rm {CNOT}_{\phi_q=\zeta}
=  {\rm \mathit{I}_0 \otimes Z_{\zeta}} \cdot \rm {CNOT}_{\phi_q=0}.
\end{gather}

This implies that we only need to implement the complete exchange between the states $|11\rangle$ and $|02\rangle$ of the $\sqrt{\rm CZ}$ operation, 
regardless of the specific transition path.
The path-independent property contributes to simplifying waveform model, and further reduce leakage under certain parameters ~\cite{watanabe2024zz,yi2024robust,hyyppa2024reducing}.

\begin{figure}[ht]
\centering
\includegraphics[width=1.0\linewidth]{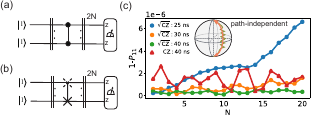}
\caption{\justifying
    (a) and (b) depict the quantum circuit diagrams used to test the leakage of $\rm CZ$ and $\sqrt{\rm CZ}$ gates, respectively. 
    Both operations are realized by using flat-top Gaussian waveform.
    (c) Comparison of leakage under operations with different durations.
}
\label{fig S path error}
\end{figure}

We implemented simulations comparing the optimized $\sqrt{\rm CZ}$ gates with standard CZ gates in terms of leakage without decoherence within a qubit-coupler-qubit system,
while Figs.~\ref{fig S path error} (a) and (b) show experimental setups.
We excited the qubit pair to state $|11\rangle$ and repeatedly applied optimized ${\rm CZ}$ or $\sqrt{\rm CZ}$ gates to observe the leakage under different numbers of operations.
Observations indicate that both ${\rm CZ}$ and $\sqrt{\rm CZ}$ gate exhibit leakage at different levels, ranging from $10^{-5}$ to $10^{-6}$.
Short-duration $\sqrt{\rm CZ}$ operations may also lead to cumulative errors due to imperfect exchange processes.
With sufficiently long durations, $\sqrt{\rm CZ}$ gate exhibits better leakage performance compared to ${\rm CZ}$ gate, possibly due to a larger optimization parameter space. 
Considering effects of decoherence and distortion, we ultimately chose a duration of 30 ns for the $\sqrt{\rm CZ}$ gate.

\section{Additional Verification}{\label{Additional Verification}}

\begin{figure}[ht]
\centering
\includegraphics[width=1.0\linewidth]{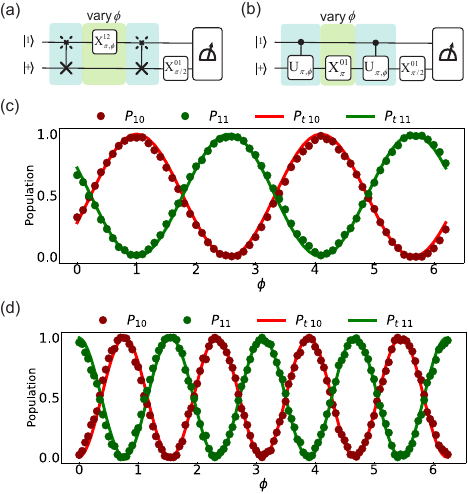}
\caption{\justifying
    (a) Experimental setups for the direct verification of tunable phase.
    (b) is designed based on (a) with the insertion
    of an $X_{\pi}^{01}$ gate to serve as an echo, accumulating the
    controlled phase effects.
    (c) and (d) correspond to	experimental results of (a) and (b) respectively, with 5000
    shots. 
    Theoretical predictions (lines) are consistent with experimental data (dots).
}
\label{fig S tunable}
\end{figure}

This section provides additional details on the experimental validation of controlled-phase operations discussed in the main text.
We designed a Ramsey-type experiment to characterize the controlled phase of the CU operation and enhance the measurement sensitivity using an echo technique.
Experimental waveform sequences and results are depicted in Fig.~\ref{fig S tunable}.

To directly test the tunable $\phi$, the qubit pair is prepared in state $|1+\rangle$.
We conducted final-state measurement after applying CNOT opertions with varying $\phi$ (${\rm CU}_{\pi,\phi}$) followed by a layer of rotation.
Mathematically, this process is described as:
\begin{align}	
S'&={\rm \mathit{I}_0\otimes X^{01}_{\pi/2,0}} \cdot {\rm CU}_{\pi,\phi} \cdot S\\
&= 
-\frac{1}{2}e^{i\phi_q}
\begin{pmatrix}
    0 	\\
    0 	\\
    e^{-i\phi} + e^{i\phi} \\
    - ie^{-i\phi} + ie^{i\phi} 
\end{pmatrix},\\
S &= (0,0,\frac{1}{\sqrt{2}},\frac{1}{\sqrt{2}})^{T},
\end{align}
where $S$ ($S'$) corresponds to the initial (final) state vector, $\phi$ represents the controlled phase, and $\phi_0$ is the adjustable single-qubit phase.
Hence, theoretically, the population of final states are:

\begin{gather}
P_{10} = \cos(\phi')^2  = \cos(\phi+\phi_0)^2, \label{P10}\\
P_{11} = \sin(\phi')^2 = \sin(\phi+\phi_0)^2. \label{P11}
\end{gather}
$P_{10}$ and $P_{11}$ follow a sinusoidal oscillation with a period of $\pi$ when $\phi \in [0,2\pi]$, 
which agree with Eqs.~\eqref{P10} and ~\eqref{P11}.

We further introduced an additional $X^{01}_{\pi,0}$ gate between two ${\rm CU}_{\pi,\phi}$ gates, which is described as follows:
\begin{align}	
S_{\rm echo}'&={\rm \mathit{I}_0 \otimes X^{01}_{\pi/2,0}} \cdot {\rm CU}_{\pi,\phi} \cdot {\rm \mathit{I}_0 \otimes X^{01}_{\pi,0}} \cdot {\rm CU}_{\pi,\phi} \cdot S\\
&= \frac{1}{2}e^{2i(\phi_q-\phi)} 		
\begin{pmatrix}
    0 	\\
    0 	\\
    i(1+e^{2i\phi}) \\
    (1-e^{2i\phi})
\end{pmatrix},
\end{align}
and the population of final states are as follows:
\begin{gather}
P_{10} = \cos(2\phi')^2 = \cos(2(\phi+\phi_0))^2,\\
P_{11} = \sin(2\phi')^2 = \sin(2(\phi+\phi_0))^2.
\end{gather}

The $\rm X^{01}_{\pi,0}$ gate serves as an echo during the evolution, doubling the influence of $\phi$.
The echo causes oscillations with the period $T = \pi/2$,
as shown in the experiment and simulation data.
These results are consistent with our theoretical expectations, 
implying our precise control over the controlled phase.

\section{Experimental Quantum Circuits}{\label{Quantum Circuits}}

\begin{figure}[H]
\centering
\includegraphics[width=1.0\linewidth]{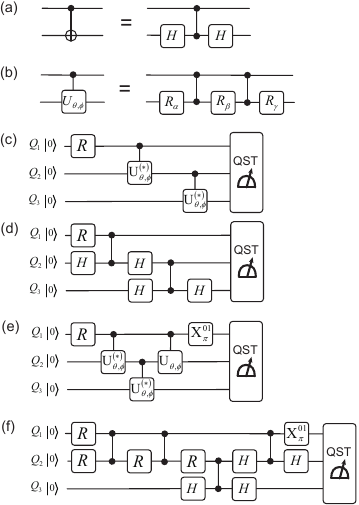}
\caption{\justifying
    The diagrams of quantum circuit used in the experiment. 
    (a) CZ-based CNOT gate.
    (b) CZ-based CU gate.
    $R_{i}$ ($i=\alpha,\beta,\gamma$) is the ${\rm SU_2}$ single-qubit gate composed of two $\rm X_{\pi/2}^{01}$ gates, treated as a single operation here for consistency.
    (c) The quantum circuits based on TCG scheme for preparing the three-qubit GHZ state and (e) the three-qubit W state.
    All two-qubit gates are implemented using TCG-based CU gates. 
    (d) and (f) are traditional quantum circuits for GHZ state and W state, respectively.
}
\label{fig S circuit}
\end{figure}

\bibliography{maintext}


\nocite{*}

\end{document}